\documentclass[prl,nofootinbib,showpacs,preprint]{revtex4-1}

\usepackage{subfigure}
\usepackage{amsmath}
\usepackage{amsfonts}
\usepackage{amssymb}
\usepackage[dvips]{graphicx}

\usepackage[colorlinks]{hyperref}
\usepackage[usenames,dvipsnames]{color}

\usepackage[applemac]{inputenc}
\usepackage[T1]{fontenc}
\usepackage{amssymb}
\usepackage{amsfonts}
\usepackage{subfigure}
\usepackage{lipsum}
\usepackage{amsfonts} 
\usepackage{braket}
\usepackage{graphicx} 
\usepackage[usenames]{color} 

\usepackage{bbm}
\def\beq{\begin{equation}}
\def\eeq{\end{equation}}
\def\bsp{\begin{split}}
\def\esp{\end{split}}
\def\bea{\begin{eqnarray}}
\def\eea{\end{eqnarray}}
\def\ba{\begin{array}}
\def\ea{\end{array}}

\def\l.{\left.}
\def\r.{\right.}

\def\part{\partial}

\begin{document}

\preprint{UdeM-GPP-TH-16-250}
\preprint{arXiv:16xx.xxxx}
\title{Graviton laser}
\author{A. Landry} 
\email{alexandre.landry.1@umontreal.ca}
\author{M. B. Paranjape} 
\email{paranj@lps.umontreal.ca}
\affiliation{Groupe de physique des particules, D\'epartement de physique,
Universit\'e de Montr\'eal,
C.P. 6128, succ. centre-ville, Montr\'eal, 
Qu\'ebec, Canada, H3C 3J7 }

\begin{abstract}

\section{Abstract}  
We consider the possibility of creating a graviton laser.  The lasing medium would be a system of contained, ultra cold neutrons.  Ultra cold neutrons are a quantum mechanical system that interacts with gravitational fields and with the phonons of the container walls.  It is possible to create a population inversion by pumping the system using the phonons.  We compute the rate of spontaneous emission of gravitons and the rate of the subsequent stimulated  emission of gravitons.   The gain obtainable is directly proportional to the density of the lasing medium and the fraction of the population inversion.  The applications of a graviton laser would be interesting.

\vskip 1in
\leftline{“Essay written for the Gravity Research Foundation 2016 Awards for Essays on Gravitation.”}

\leftline{Corresponding author M. B. Paranjape, paranj@lps.umontreal.ca}
\end{abstract}


\maketitle


\section{Introduction}
A laser works by the principle that in a (lasing) medium, where there are more quantum excited states than de-excited states, spontaneously emitted photons  can subsequently stimulate the emission of more photons that are coherent with the stimulating photon.  The process can cascade multiple times creating an avalanche of coherent photons.  If the gain of the lasing medium is sufficiently high, a significant amplification can be obtained with a single pass, there is no need for reflecting mirrors.  In this essay, we examine the possibility that such amplification can be obtained for gravitons.  

Ultra-cold neutrons, \cite{golub1991ultra} are normally defined to have such low kinetic energy that they are unable to penetrate the material walls of a container.  The kinetic energy of the neutrons has to be typically less than about 300 nano eV.  When the energy of a neutron is in this range, then its de Broglie wavelength is of the order of 100 nanometres, which means that the neutron coherently senses at once almost a thousand nuclei of the wall of the container, actually through the strong interaction.  The neutrons have negligible other interactions with the material of the container.  The zero Fourier component of the strong interaction between the nuclei and the neutron is all that is relevant and corresponds to a constant energy valued work function.  To be able to penetrate into  into the walls of the container, the neutrons have to have greater kinetic energy than the threshold.  Failing that, the neutrons are reflected perfectly.  We will consider an ensemble of even colder neutrons.  Mili-Kelvin neutrons have kinetic energies that are of the order of a few pico eV.  At such low energies, the neutrons become sensitive to the gravitational field of the earth.  Indeed the Q-bounce experiment \cite{Jenke2009318} measures exactly the consequences the quantum mechanical interaction of the neutrons with this gravitational field.  

For mili-Kelvin neutrons bouncing on a smooth, horizontal base, the Schrödinger equation for the neutrons is easily, analytically solved.  The potential in the $z$ direction is $m_N g z$  (where $m_N$ is the neutron mass and $g$ is the earth's gravitational acceleration) and the boundary condition at $z=0$ corresponds to an effectively, infinite energy barrier.  The eigenfunctions are simply the Airy functions $\psi_n(z)=Ai(\frac{z}{z_0}-\alpha_n)$ where $z_0^3=\frac{\hbar^2}{2gm_N^2}$  and $-\alpha_n$ is the $n$th zero of the Airy function with energy eigenvalue $E_n=m_Ngz_0\alpha_n$ \cite{Landry:2016zip}.  An approximate analytic formula for the energy levels is given by $E_n\approx 1,69 \left(n-\frac{1}{4}\right)^{2/3}peV$.  Transitions between these quantum levels have already been induced and observed by  pumping the system by mechanically vibrating the base at the appropriate frequency, approximately 600Hz.  The neutrons can be lifted to say the third excited level using the mechanical pumping method. Pumping at the resonant frequency for the transition between the first and third levels asymptotically will populate each of these levels equally. But then there will be a population inversion with respect to the second excited level.  Thus spontaneous emission of a graviton from a transition from the third to the second level, and the subsequent stimulated emission of gravitons could yield significant, laser type amplification of the gravitational wave.

\section{Quantum gravitons and the amplitude for spontaneous and stimulated emission}
In a zero order approximation, gravitons correspond to the non-interacting excitations of the quantized, linearized Einstein equations.  These equations describe the free field dynamics of a massless spin 2 field.  The quantization is trivial and well understood.  The quantum gravitational field corresponds to the tensor $h_{\mu\nu}$, which can be decomposed into its Fourier modes
\beq
h_{\mu\nu}=\int d^3k \left(\frac{e^{-ik_\mu x^\mu}}{\sqrt V}A(\vec k)\epsilon_{\mu\nu}+\frac{e^{ik_\mu x^\mu}}{\sqrt V}\epsilon^*_{\mu\nu}A^\dagger(\vec k)\right)
\eeq
where $\epsilon_{\mu\nu}$ is the polarization tensor for massless spin two particles and the operators $A^\dagger(\vec k)$  and $A(\vec k)$ satisfy the simple algebra of annihilation and creation operators
\beq
\left[A(\vec k),A^\dagger(\vec k')\right]=\delta^3(\vec k-\vec k').
\eeq
These operators respectively create and destroy one free graviton of momentum $\vec k$.  The frequency satisfies $k_0=|\vec k|$ which is appropriate for a massless particle.  

The gravitational field of the graviton, interacts with the system of ultra-cold neutrons by distorting the space-time in which the neutron is immersed.  The ambient space-time is that which corresponds to the gravitational field of the earth.  This space-time is simply the Schwarzschild geometry
\beq
c^2d\tau^2=\left(1-\frac{2GM_\oplus}{c^2 r}\right)c^2dt^2-\left(1-\frac{2GM_\oplus}{c^2 r}\right)^{-1}dr^2-r^2d\Omega^2
\eeq
where the symbol $\oplus$ stands for the earth.  Superimposed on this space-time is that of the gravitational wave of the graviton
\beq
c^2d\tau^2=c^2dt^2-dx^2-(1+h(x-ct))dy^2-(1-h(x-ct))dz^2
\eeq
for a gravitational wave propagating in the $x$ direction polarized in the + sense in Minkowski space-time.  In the weak field linear approximation, the combined metric is
\bea
\nonumber c^2d\tau^2=\left(1-\frac{2GM_\oplus}{c^2r}\right)c^2dt^2-\left(1+\frac{2GM_\oplus}{c^2r}\right)(\vec x\cdot \vec dx)^2/r^2\\-h(x-ct)dy^2+h(x-ct)dz^2-r^2d\Omega^2.
\eea
The classical dynamics of the neutrons, apart from the reflection from the base, is governed by the geodesic equation:
\beq
\frac{d^2x^i}{d\tau^2}+\Gamma^i_{\mu\nu}\frac{dx^\mu}{d\tau}\frac{dx^\nu}{d\tau}=0
\eeq
The connection is given to first order by $\Gamma^\lambda_{\mu\nu}=(1/2)\eta^{\lambda\sigma}(\partial_\mu h_{\sigma\nu}+\partial_\nu h_{\sigma\mu}-\partial_\sigma h_{\mu\nu})$.
We follow closely the exposition by Bertschinger in \cite{bertschinger}, decomposing the gravitational field as
\beq
h_{00}=-2\phi,\quad h_{0i}=w_i,\quad h_{ij}=-2\psi\delta_{ij}+2s_{ij}
\eeq
where $\delta^{ij}s_{ij}=0$, we have $w_i=0$, $\phi=GM_\oplus/r$, $\psi=GM_\oplus/3r$ and $s_{ij}=s^{\oplus}_{ij}+s^{\approx}_{ij}$ with $s^{\oplus}_{ij}=(2GM_\oplus/r^3)(x_ix_j-(r^2/3)\delta_{ij})$ and  $s^{\approx}_{zz}=-s^{\approx}_{yy}=h(x-t)$ where we now add the use the symbol $\approx$ for the gravitational wave.

The Hamiltonian for the system is given by
\beq
H(x^i,\pi_j)=(1+\phi)E(p_j)\,\,{\rm where}\,\, E(p_j)=(\delta^{ij}p_ip_j+m_N^2)^{1/2}
\eeq
with $p_i=(1+\psi)\pi_i-(\delta^{ij}\pi_i\pi_j+m_N^2)^{1/2}w_i-s^j_{\,\,\, i}\pi_j$ which in our case becomes just $p_i=(1+\psi)\pi_i-s^j_{\,\,\, i}\pi_j$.  As $\pi_i$ and $p_i$ differ only by first order terms, we get the Hamiltonian to first order
\beq
H(x^i,\pi_j)=E(\pi_j)+\frac{(\psi\delta^{ij}-s^{ij})\pi_i\pi_j}{E(\pi_j)}+\phi E(\pi_j).
\eeq 
Expanding to second order in the canonical momentum gives
\beq
H(x^i,\pi_j)=m_N+\frac{|\vec\pi|^2}{2m_N}+\frac{(\psi\delta^{ij}-s^{ij})\pi_i\pi_j}{m_N}+\phi (m_N+\frac{|\vec\pi|^2}{2m_N}).
\eeq
From this expression we extract our basic Hamiltonian of the neutron interacting with the gravitational field of the earth
\beq
H_0(x^i,\pi_j)=\frac{|\pi|^2}{2m_N}+m_N\phi\label{hfree}
\eeq
subtracting off the rest mass, and the perturbation, which we split into two terms
\bea
H^\oplus(x^i,\pi_j)&=& \frac{\phi|\vec\pi|^2}{2m_N}+\frac{(\psi\delta^{ij}-s^{\oplus\,ij})\pi _i\pi_j}{m_N}\label{hearth}\\
H^\approx(x^i,\pi_j)&=&\frac{-s^{\approx\, ij}\pi _i\pi_j}{m_N}.\label{hwave}
\eea

The Schrödinger equation resulting from Eqn. \eqref{hfree} is essentially free in the $x$ and $y$ directions, but in the $z$ direction it is
\beq
\left(\frac{-\hbar^2}{2m_N}\frac{d^2}{dz^2} +m_Ngz\right)\psi(z,t)=E\psi(z,t).\label{schrodinger}
\eeq
the perturbation from Eqn.\eqref{hearth} induces some perturbative changes in the energies and wavefunctions of $H_0(x^i,\pi_j)$, however, these changes are static and do not give rise to any transitions, spontaneous or stimulated between levels.  The time dependent perturbation that is relevant, from Eqn.\eqref{hwave}  (replacing $h(x-ct)\rightarrow \frac{\sqrt{G}}{c^2}h(x-ct)$ the canonically normalized metric perturbation) is
\beq
H^\approx(x^i,\pi_j)=- \frac{\sqrt{G}}{c^2}\frac{h(x-ct)\pi_z^2}{m_N}.
\eeq
Then amplitude for spontaneous or stimulated emission of a graviton is proportional to the matrix element  $\langle \psi_{n'}|-\frac{\pi_z^2}{m_N}|\psi_n\rangle $ which, using Eqn.\eqref{schrodinger}, gives
\bea\nonumber
\langle \psi_{n'}|-\frac{\pi_z^2}{m_N}|\psi_n\rangle &=& \langle \psi_{n'}|(2m_N g z -2E_n)|\psi_n\rangle\\
&=&\langle \psi_{n'}|2m_N g z |\psi_n\rangle.
\eea
The matrix element is easily computed using the exact eigenfunctions and the integral
\beq
\int_{0}^{\infty} dy y  Ai(y -\alpha_{m} ) Ai(y -\alpha_{n})=\frac{-2Ai'(-\alpha_m)Ai'(-\alpha_n)}{(\alpha_m-\alpha_n)^2}
\eeq
which gives
\beq
\langle \psi_{n'}|2m_N g z |\psi_n\rangle=-\frac{4 m_N g z_0}{(\alpha_{n'}-\alpha_n)^2}.
\eeq
The rate, per unit time and volume, of spontaneous or stimulated emission of gravitons is then is given by, modifying a calculation in Baym, \cite{gordon1990lectures} 
\bea
d\Gamma^{\rm emm.}_{n'\rightarrow n}&=&\frac{G}{c^4}\frac{4\pi^2c^2}{ck V}(N_k +1)\times\nonumber\\
&\times&|\langle \psi_{n'}|2m_N g z |\psi_n\rangle|^2\delta(E_{n'}-E_n-\hbar c k)\label{infrate}
\eea
where $N_k$ is the number of photons involved in the stimulated emission, and when $N_k=0$ we get the rate for spontaneous emission.  (The whole system is imagined in a box of volume $V$.)  There are 
$ V {d^3k}/{(2\pi)^3}$ states in the volume $d^3k$ of phase space and as  $E_k=\hbar c |k|$
\beq
V\frac{d^3k}{(2\pi)^3}=V E_k^2\frac{dE_kd\Omega}{(2\pi\hbar c)^3}
\eeq 
states per unit energy. Then integrating Eqn.\eqref{infrate} over the energy of the emitted graviton we get
\beq
d\Gamma^{\rm emm.}_{n'\rightarrow n}=\frac{G}{c^4}\frac{(E_{n'}-E_n)d\Omega}{ 2 \pi\hbar^2 c}(N_k +1)\left(\frac{4 m_N g z_0}{(\alpha_{n'}-\alpha_n)^2}\right)^2\label{rate1}
\eeq
Since $E_n=m_N g z_0\alpha_n$ where $z_0^3=\frac{\hbar^2}{2gm_N^2}$ and integrating over directions, we get
\bea
\Gamma^{\rm emm.}_{n'\rightarrow n}&=&\frac{2G}{\hbar^2c^5}(N_k +1)\left(\frac{ m_N g z_0}{\alpha_{n'}-\alpha_n}\right)^3\nonumber\\
&=&\frac{G}{\hbar^2c^5}(N_k +1)\left(\frac{ m_N g^2 \hbar^2}{(\alpha_{n'}-\alpha_n)^3}\right)   \nonumber\\
&=&\frac{G}{c^5}(N_k +1)\left(\frac{ m_N g^2 }{(\alpha_{n'}-\alpha_n)^3}\right)       \label{rate}
\eea
\section{Population Inversion}
We can obtain a population inversion by mechanically vibrating the base at the resonant frequency corresponding to the transition between two levels \cite{Jenke2009318}.  Resonant pumping of a system has the property that the probability of inducing a transition from level $n$ to level $n'$ is equal to the probability of inducing a transition in the opposite direction. The long time behaviour of such a system is governed by the simple equations, with ${\cal N}_n(t)$ the occupation number of the neutrons in the $n$th level,
\bea
\dot {\cal N}_{n'}(t)&=&-\lambda  {\cal N}_{n'}(t)+\lambda  {\cal N}_{n}(t)\nonumber\\
\dot {\cal N}_{n}(t)&=&-\lambda  {\cal N}_{n}(t)+\lambda  {\cal N}_{n'}(t)
\eea
where $\lambda$ is the effective rate of transitions per unit time.   The solution of this system is
\bea
{\cal N}_{n'}(t)&=&\frac{({\cal N}_{n'}(0)(1+e^{-2\lambda t})+{\cal N}_n(0)(1-e^{-2\lambda t}))}{2}\\
{\cal N}_{n}(t)&=&\frac{({\cal N}_{n}(0)(1+e^{-2\lambda t})+{\cal N}_{n'}(0)(1-e^{-2\lambda t}))}{2}
\eea
Hence in the long time limit, the populations go to ${\cal N}_{n'}(t)={\cal N}_n(t)\rightarrow\frac{1}{2}({\cal N}_{n}(0)+{\cal N}_{n'}(0))$.

If we pump the system between levels $n$ and $n+m$, with $m>1$, there will be intermediate levels in between the two with respect to which there will be a population inversion. Then one could imagine spontaneous emission of a graviton by a transition from the upper level to one of the intermediate levels.

The amplification factor $\kappa$ of a lasing medium is given by
\beq
\kappa=\sigma ({\cal N}_{n'}-{\cal N}_n)
\eeq
where $\sigma$ is the cross section of induced transitions between the levels of the medium, \cite{tarasov2014laser}.  The cross section is trivially related to the transition rate for spontaneous emission computed above in Eqn.\eqref{rate} by dividing by the incident flux.  Evidently a population inversion, ${\cal N}_{n'}>{\cal N}_n$ is necessary for any amplification.
\section{Discussion}
The rate computed in Eqn.\eqref{rate} is absurdly small for terrestrial applications.
  But if the neutron is  in a highly excited state we do get some enhancement.  The energy levels behave as $E_n\sim n^{2/3}$, which comes from a similar behaviour of the zeros of the Airy functions.  Clearly for two highly excited levels $n+m$ and $n$ as $n\to\infty$, we get 
\beq
\alpha_{n+m}-\alpha_{n}\sim \frac{m}{n^{1/3}}.
\eeq 
Therefore the denominator in Eqn.\eqref{rate} can in principle be made arbitrarily small, giving a corresponding enhancement of the probability of spontaneous emission.  In general, for levels separated by gravitons of energies in the deep infrared limit, there is no suppression.  Observing and controlling stimulated amplification of gravitons would be of unimaginable importance.

One can also imagine the process of lasing amplification occurring on the surface of a star, say a neutron star.  Here the surface gravity can be $10^{12}$ times greater than on the surface of the earth.  This yields an additional factor of $10^{24}$.  Coupled with a pair of excited levels, it is not impossible that there could be a lasing amplification of spontaneously   emitted gravitons.  Furthermore, speculatively, if $m_N$ is replaced by a considerably heavier, as yet undiscovered, supersymmetric or dark matter particle, again there could be substantial increase in the rate of spontaneous emmission.  Finally, one can imagine, in an astrophysical scenario, that an enormous number of gravitons are somehow released in some cataclysmic event, for example, the recently observed black-hole merger event, \cite{Abbott:2016blz}.  Then $N_k$ could be sufficiently high to compensate for the other  factors that are rendering the amplitude small.

\section{Acknowledgments} We thank Biothermica Corporation, Sibylla Hesse foundation and NSERC of Canada for their financial support.  We thank Bhujyo Bhattacharya and Eric Dupuis for useful discussions. 
\bibliographystyle{apsrev}
\bibliography{ref}

\end{document}